# Compressive Sensing Imaging of 3-D Object by a Holographic Algorithm

Shiyong Li, *Member, IEEE*, Guoqiang Zhao, Houjun Sun, and Moeness Amin, *Fellow, IEEE*

*Abstract*—Existing three-dimensional (3-D) compressive sensing-based millimeter-wave (MMW) imaging methods require a large-scale storage of the sensing matrix and immense computations owing to the high dimension matrix-vector model employed in the optimization. To overcome this shortcoming, we propose an efficient compressive sensing (CS) method based on a holographic algorithm for near-field 3-D MMW imaging. An interpolation-free holographic imaging algorithm is developed and used as a sensing operator, in lieu of the nominal sensing matrix typically used in the CS iterative optimization procedure. In so doing, the problem induced by the large-scale sensing matrix is avoided. With no interpolations required, both the computational speed and the image quality can be improved. Simulation and experimental results are provided to demonstrate the performance of the proposed method in comparison with those of the *ωK* based CS and the traditional Fourier-based imaging techniques.

## I. INTRODUCTION

MILLIMETER-WAVE (MMW) has attractive characteristics compared with waves in the microwave band or lower frequency bands. These include higher carrier frequency and wider usable frequency band that enables higher target cross-range and down-range resolutions. Another important feature of MMW is the design of small and light systems and equipment. Accordingly, it is beneficial to adopt MMW for short-range broadband communications [1], [2], high-resolution sensing [3], [4], and radio astronomy [5]. MMW imaging techniques have been widely developed and applied to non-destructive testing [6], biomedical diagnosis [7], and personnel security inspection [8]-[11].

MMWs are capable of penetrating regular clothing to form an image of a person and concealed objects. Most importantly, MMWs are non-ionizing and, therefore, pose no known health hazard at moderate power levels. MMW imaging systems can be classified into two types: passive imaging systems and active imaging systems. Both types have their own offerings and challenges. This paper focuses on active MMW imaging techniques, which typically implement large scale antenna arrays to illuminate the whole human body, leading to high system cost. Compressive sensing (CS) and sparse reconstruction techniques, on the other hand, have been used to reduce the number of array elements, thereby the system cost, without degradation of image quality [12]-[14]. CS has been extensively studied in radar imaging [15]-[19], microwave imaging [20]-[23], array synthesis and diagnosis [24], [25], and direction-of-arrival estimation [26]-[28].

CS methods are typically based on the matrix-vector model, which employs a large-scale sensing matrix in 3-D MMW imaging for personnel inspection and security applications. Such model was incorporated in [29] for a 3-D compressive phased array imaging. CS was applied to single-frequency 2-D MMW holographic imaging in [30] and [31], where a Fourier-based imaging operator represented the sensing matrix. This replacement, in essence, avoided storing and processing of a large-scale sensing matrix which, in turn, simplified imaging and permitted its realization on an ordinary personal computer. In [32], the 3-D *ωK* algorithm, referred to as range migration algorithm, was used in combination with the CS principle for image reconstruction. Nevertheless, this algorithm includes the forward and inverse Stolt interpolations that entail require high computations and can lead to reduced image fidelity. An interpolation-free SAR imaging algorithm, named range stacking, was proposed in [33] and extended to 3-D imaging in [34]. The range stacking reconstruction method forms the target image at different range points by matched filtering the SAR signal in the spatial frequency domain. The result is integrated over frequencies to yield the marginal Fourier transform of the target function.

In [35] and [36], we considered a single-frequency based auto-focus holographic imaging algorithm. The auto-focus was obtained by calculating the amplitude integral values of the images reconstructed at different focusing range bins. These values draw to a minimum when the image is well focused. In this paper, we extend the above algorithm to the wideband signal, and change the integral variables of the imaging model to eliminate the Stolt interpolation. This changing is similar to the work in [33], [34]. However, unlike these references, we first apply the inverse Fourier transform over the azimuth- and elevation-frequencies with respect to the matched filtered data.

Manuscript received February 16, 2018. This work was supported by the National Natural Science Foundation of China under Grant 61771049.

S. Li, G. Zhao and H. Sun are with the Beijing Key Laboratory of Millimeter Wave and Terahertz Technology, Beijing Institute of Technology, Beijing 100081, China. S. Li is also with the Center for Advanced Communications, Villanova University, Villanova, PA 19085 USA. (e-mail: lisy_98@bit.edu.cn).

M. Amin is with the Center for Advanced Communications, Villanova University, Villanova, PA 19085 USA (e-mail: moeness.amin@villanova.edu).

The results are then integrated over the fast-time frequencies, which can be considered as a coherent summation of the single-frequency holographic imaging results. These steps amount to a linear relationship between the scene and the measurements, defined by a sensing operator, and as such, the CS problem can be readily formulated. Due to the fact that interpolations are avoided in the iterations underlying the optimization algorithm, the computational speed and optimization solution can both be improved in comparison with the 3-D $\omega K$-based CS method [32].

In order to deal with compressed data, which corresponds to selecting few antennas, we utilize a uniform-random spatial undersampling scheme, in lieu of the totally random undersampling scheme. The benefits of the former over the latter sampling strategy can be made evident by analyzing the mutual coherence measure of the sensing operator. In imaging, this measure, in essence, represents the maximum sidelobe value of the point spread function (PSF) in Fourier-based imaging. Accordingly, it can be used to quantify the effect of different undersampling schemes on performance. Also, we compare the PSFs of the proposed sensing operator and that of the $\omega K$-based sensing operator, and show the superior performance of the former.

The remainder of this paper is organized as follows. Section II presents the formulation of the 3-D interpolation-free holographic imaging algorithm, which is used to construct the operator considered as the sensing matrix of CS. In Section III, we provide the CS imaging method and present the uniform-random under-sampling scheme. The relationship between the mutual coherence and the point spread function is analyzed. Numerical simulations and experimental results are shown in Section IV. Finally, concluding remarks are presented in Section V.

## II. 3-D INTERPOLATION-FREE HOLOGRAPHIC IMAGING ALGORITHM

### A. Formulation of the interpolation-free holographic imaging algorithm

The configuration of the imaging system is shown in Fig. 1. For a frequency modulated continuous wave (FMCW) transceiver, the transmitted signal can be expressed as,

$$s_T(t) = \exp\left[j2\pi\left(f_0 t + \frac{1}{2}Kt^2\right)\right], \quad (1)$$

where $f_0$ is the center frequency, $t$ is the fast-time variable within one pulse repetition interval, and $K$ is the frequency slope of the transmitted signal. The backscattered signal from a point target is given by,

$$s_R(x', y', t) = \sigma(x, y, z) s_T(t - \tau_d), \quad (2)$$

where $\sigma(x, y, z)$ represents the backscattering coefficient of the point target at location $(x, y, z)$, and $\tau_d$ is the round-trip time-delay defined by the propagation speed of the electromagnetic wave and the distance from the receiver at $(x', y', Z)$ to the target. All array elements lie on plane $Z$.

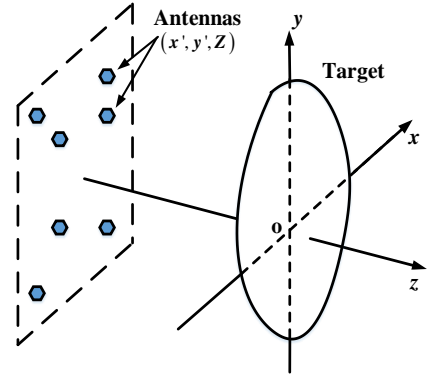

Fig. 1. Geometrical configuration of the imaging system.

Dechirp-on-receive is used to demodulate the received signal, and yields the intermediate frequency signal,

$$s_{IF}(x', y', f) = \sigma(x, y, z) \exp\left[-j2\pi(f + f_0)\tau_d\right], \quad (3)$$

where $f = Kt$ is considered as the fast-time frequency. We assume the residual video phase (RVP) has been removed. The time-delay is given by,

$$\tau_d = \frac{2\sqrt{(x-x')^2 + (y-y')^2 + (z-Z)^2}}{c}.$$

Due to the target located in the near-field, which is illuminated by spherical waves, the square root of the above time-delay expression cannot be simplified, as in the case of far-field. For a volume target, extending in all three dimensions,

$$s_{IF}(x', y', f) = \int \sigma(x, y, z) e^{-j2k\sqrt{(x-x')^2 + (y-y')^2 + (z-Z)^2}} dx dy dz, \quad (4)$$

where $k = 2\pi(f + f_0)/c$ is considered as the wavenumber. In the above equation, we have ignored the propagation loss of spherical waves which is characterized by $1/\left[(x'-x)^2 + (y'-y)^2 + (Z-x)^2\right]$ for the round-trip propagation. The exponential term in (4) represents a spherical wave emanating from $(x', y', Z)$. This term can be decomposed into a superposition of plane wave components [8], as

$$e^{-j2k\sqrt{(x-x')^2+(y-y')^2+(z-Z)^2}} = \iint e^{-jk_{x'}(x-x')} e^{-jk_{y'}(y-y')} e^{-jk_z(z-Z)} dk_{x'} dk_{y'}, \quad (5)$$

where $k_{x'}$ and $k_{y'}$ are the Fourier transform variables corresponding to $x'$ and $y'$, respectively. The spatial frequencies $k_{x'}$ and $k_{y'}$ range from $-2k$ to $2k$. Substituting (5) into (4), and using the Fourier transform, we obtain [8],

$$\sigma(x, y, z) = \iiint S_{IF}(k_x, k_y, k) e^{-jk_z Z} e^{jk_x x} e^{jk_y y} e^{jk_z z} dk_x dk_y dk_z, \quad (6)$$

where $k_z = \sqrt{4k^2 - k_x^2 - k_y^2}$, $S_{IF}(k_x, k_y, k) = \text{FFT}_{x,y}\{s_{IF}(x, y, f)\}$ and $\text{FFT}_{x,y}\{\cdot\}$ indicates the 2-D fast Fourier transform (FFT) over $(x, y)$. The distinction between the primed and unprimed coordinates is now dropped since the two coordinates coinciding. Typically, the data is sampled at uniform intervals of position $(x, y)$ and fast-time frequency $f$ or $k$. These samples are nonuniformly spaced in $k_z$ and, as such, require resampling at equally spaced positions which is referred to as Stolt interpolation [37]. To avoid this interpolation, and similar to the approach in [33], [34], we change the integral variable $dk_z$ in (6) to $dk$ and modify the integrant as follows:

$$\sigma(x, y, z) = \iiint S_{IF}(k_x, k_y, k) e^{j\sqrt{4k^2 - k_x^2 - k_y^2}(z-Z)} e^{jk_x x} e^{jk_y y} J(k, k_x, k_y) dk_x dk_y dk, \quad (7)$$

where $J(k, k_x, k_y)$ is given by,

$$J(k, k_x, k_y) = \frac{\partial k_z}{\partial k} = \frac{4k}{\sqrt{4k^2 - k_x^2 - k_y^2}}.$$

In [33], [34], the summation over fast-time frequencies was carried out prior to performing the inverse FFT over the spatial frequencies, with matched filter used to specify the particular range bin. On the opposite, in this paper, we first perform the inverse FFT with respect to the matched filtered outputs over the spatial frequencies $k_x$ and $k_y$, then the results are integrated over all fast-time frequencies (or wavenumbers). In this way, imaging can be considered as an extension of 2-D single-frequency holographic imaging algorithm.

We set $S_{IF}'(k_x, k_y, k; z) = S_{IF}(k_x, k_y, k) e^{j\sqrt{4k^2 - k_x^2 - k_y^2}(z-Z)} J(k, k_x, k_y)$, and substituting it into (7) yields,

$$\sigma(x, y, z) = \int \left\{ \iint S_{IF}'(k_x, k_y, k; z) e^{jk_x x} e^{jk_y y} dk_x dk_y \right\} dk. \quad (8)$$

Clearly, the inner double integral represents a 2-D inverse Fourier transform. Thus,

$$\sigma(x, y, z) = \int \text{IFFT}_{(k_x, k_y)} \left\{ S_{IF}'(k_x, k_y, k; z) \right\} dk, \quad (9)$$

where $\text{IFFT}_{(k_x, k_y)} \{\cdot\}$ represents the 2-D inverse FFT over $(k_x, k_y)$. The above integral over $k$ can be performed in parallel for all range bins, as also stated in [33] and [34]. The above steps eliminate Stolt interpolations.

It is noted from (7) to (9) that the inner integral over $k$ is identical to the single-frequency holographic imaging algorithm at a specified range bin [8]. Therefore, the above 3-D imaging is accomplished by the following procedure: We first perform a 2-D single-frequency based imaging in parallel at different specified range bins with respect to all individual fast-time frequencies. Then, these images are summed coherently over frequencies, and a well-focused 3-D image can be obtained. Therefore, the integral (9) can be considered as an extension of single-frequency holographic imaging algorithm, and is referred to as 3-D holographic imaging algorithm, for convenience. A byproduct of the above problem formulation is that we can freely choose the imaging zone of interest along the $z$ direction. This is different from the traditional 3-D $\omega K$ algorithm, which generates the image as follows [8]:

$$\sigma(x, y, z) = \text{IFFT}_{k_x, k_y, k_z} \left\{ \underset{\sqrt{4k^2 - k_x^2 - k_y^2} \to k_z}{\text{StoltInterp}} \left\{ \text{FFT}_{x, y} \left\{ s_{IF}(x, y, k) \right\} e^{-j\sqrt{4k^2 - k_x^2 - k_y^2} Z} \right\} \right\}. \quad (10)$$

The $\underset{\sqrt{4k^2 - k_x^2 - k_y^2} \to k_z}{\text{StoltInterp}} \{\cdot\}$ indicates the Stolt interpolation from $(k, k_x, k_y)$ to $(k_z, k_x, k_y)$ which is necessary for performing the 3-D inverse FFT. Due to the properties of FFT, imaging along the entire unambiguous range along the $z$ direction must be implemented.

### B. Forward and backward operators

Below, we construct a forward operator $\boldsymbol{\Phi}$ which will be used as the sensing matrix for the sparse reconstruction presented in the next section. According to (9), the backward operator $\boldsymbol{\Phi}^\dagger$ is as follows,

$$\boldsymbol{\Phi}^\dagger \{\mathbf{S}\} = \sum_k \text{IFT}_{(k_x, k_y)} \left\{ \text{FT}_{(x, y)} \{\mathbf{S}\} e^{j\sqrt{4k^2 - k_x^2 - k_y^2}(z-Z)} J(k, k_x, k_y) \right\}, \quad (11)$$

where the term in $\{\ \}$ stands for the object on which the operator acts. We can then formulate the imaging problem (9) as the following matrix form:

$$\mathbf{G} = \boldsymbol{\Phi}^\dagger \{\mathbf{S}\}, \quad (12)$$

where $\mathbf{G}$ is a 3-D matrix representing the reconstructed image $\sigma(x, y, z)$, with three dimensions corresponding to the $x$, $y$, and $z$ directions, respectively, in the image space. The symbol $\mathbf{S}$ represents the scattered data $s_{IF}(x, y, k)$ also in a 3-D matrix form, with three dimensions corresponding to the $x$ and $y$ directions of the antenna array, and the fast-time frequency dimension, respectively, in the scattered data space. Eq. (12) shows that the 3-D image is obtained through the backward operator $\boldsymbol{\Phi}^\dagger$ acting on the 3-D scattered data.

The forward operator $\boldsymbol{\Phi}$ can be obtained by inversing the process of (11) as,

$$\boldsymbol{\Phi}\{\mathbf{G}\} = \sum_z \text{IFT}_{(k_x, k_y)} \left\{ \text{FT}_{(x, y)} \{\mathbf{G}\} e^{-j\sqrt{4k^2 - k_x^2 - k_y^2}(z-Z)} J^{-1}(k, k_x, k_y) \right\}. \quad (13)$$

Clearly, no interpolations are included in (11) or (13). Similarly, $\mathbf{S} = \boldsymbol{\Phi}\{\mathbf{G}\}$.

The complexity of the $\omega K$ algorithm and the proposed algorithm is given in Table I in terms of the number of floating point operations (FLOP). Assume that the received data and the reconstructed image have a same size of $N_R \times N_A \times N_E$, along the range, azimuth, and elevation directions, respectively. We use $N_I$ to represent the interpolation kernel length which is typically chosen as 8 for the *sinc* function kernel. It is noted from Table. I that the bulk of the computation load of the two algorithms depends on the size of the image. In practice, the size of the image scene of a person usually spans $2\text{m} \times 1\text{m} \times 0.5\text{m}$ with respect to the height, width and thickness. The value of $N_R$ can assume smaller values than $N_A$ and $N_E$.

If we choose the resolution to be $5\text{mm}\times 5\text{mm}\times 3\text{cm}$ (Usually, the range resolution is lower than the transverse resolution due to the limit on bandwidth), then the computation efficiency can be computed to be $\xi = \dfrac{\text{FLOP}_{\omega K}}{\text{FLOP}_{\text{Holo3D}}} \approx 1.2$. Although this efficiency appears modest, it is proven to be important when considering the overall optimization problem, as described in the next section.

TABLE I
COMPARISON OF COMPUTATIONAL COMPLEXITY

|  | $\omega K$ algorithm | 3-D Holographic algorithm |
|---|---|---|
| Azimuth FFT | $5N_R N_E N_A \log_2 N_A$ | $5N_R N_E N_A \log_2 N_A$ |
| Elevation FFT | $5N_R N_A N_E \log_2 N_E$ | $5N_R N_A N_E \log_2 N_E$ |
| Matched filter multiplication | $6N_R N_A N_E$ | $6N_R N_A N_E$ |
| Stolt interpolation | $2(2N_i - 1)N_R N_A N_E$ | 0 |
| Azimuth IFFT | $5N_R N_E N_A \log_2 N_A$ | $5N_R N_E N_A \log_2 N_A$ |
| Elevation IFFT | $5N_R N_A N_E \log_2 N_E$ | $5N_R N_A N_E \log_2 N_E$ |
| Range IFFT | $5N_E N_A N_R \log_2 N_R$ | 0 |
| Summation along frequency | 0 | $2(N_R - 1)N_R N_A N_E$ |

## III. CS APPROACH FOR NEAR-FIELD 3-D IMAGING

In this section, we perform CS-based 3-D MMW imaging based on the aforementioned algorithm incorporating the sensing operator.

The theory of compressive sensing states that sparse signals can be recovered using far fewer samples than that required by the Nyquist sampling. Considering spatial sampling, CS can be used to reduce the number of antenna array elements, thus reducing the overall system cost. In CS, the relationship between the measurements and the sparse image assumes a linear model, namely, $\mathbf{y} = \boldsymbol{\Phi}\boldsymbol{\sigma}$, where $\mathbf{y} \in \mathbf{C}^P$ is a vector of measurement samples, $\boldsymbol{\Phi} \in \mathbf{C}^{P\times Q}$ is known as the sensing matrix with $P \leq Q$, and $\boldsymbol{\sigma}$ can be expressed as $\boldsymbol{\sigma} = \boldsymbol{\Psi}\boldsymbol{\alpha}$ where $\boldsymbol{\Psi} \in \mathbf{C}^{Q\times Q}$ is a linear sparsifying basis, and $\boldsymbol{\alpha} \in \mathbf{C}^Q$ is a sparse vector.

Due to sparsity, $\boldsymbol{\alpha}$ can be reconstructed by solving the following convex optimization problem,

$$\min_{\boldsymbol{\alpha}} \|\boldsymbol{\alpha}\|_1 \quad \text{s.t.} \quad \mathbf{y} = \boldsymbol{\Theta}\boldsymbol{\alpha}, \tag{14}$$

where $\|\boldsymbol{\alpha}\|_1 = \sum_l |\alpha_l|$ denotes the $\ell_1$ norm of $\boldsymbol{\alpha}$ and $\boldsymbol{\Theta} = \boldsymbol{\Phi}\boldsymbol{\Psi}$. Eq. (14) can be solved efficiently using several linear or quadratic programming techniques [12]. In this paper, we assume that the image is sparse in its canonical basis, and as such, set the sparsifying basis to an identity matrix.

For the 3-D MMW imaging, discussed in this paper, the elements of the sensing matrix $\boldsymbol{\Phi}$ are given by,

$$\varphi_{n,p,q} = \exp\left(-j2k_n \sqrt{(x_q - x'_p)^2 + (y_q - y'_p)^2 + (z_q - Z)^2}\right), \tag{15}$$

where $n = 1, 2, \cdots, N$, $p = 1, 2, \cdots, P$, and $q = 1, 2, \cdots, Q$. The variables $N$, $P$, and $Q$, respectively, represent the total numbers of equivalent frequencies, spatial measurement samples, and pixels of the image scene. Accordingly, the size of sensing matrix $\boldsymbol{\Phi}$ is $NP \times Q$. For the 3-D human body imaging scene, $Q$ could be of the order of $10^5$ (assuming the cross-range resolution is 1cm and the range resolution is 3cm). The number of measurements $N \times P$ could be in the order of $10^6$, under Nyquist sampling. Thus, the sensing matrix could be prohibitively large, causing challenges in both storage and processing using a personal computer. The computational complexity for one-time multiplication of this matrix with a $Q \times 1$ vector should be $6NP \times Q + 2NP(Q-1)$ FLOPs, which will be much higher than that of the operators. For a same image size and data size, as illustrated in Table I, the computation efficiency can be calculated and it is equal to $\xi = \dfrac{\text{FLOP}_{\text{Matrix}(15)}}{\text{FLOP}_{\text{Holo3D}}} \approx 5 \times 10^4$. Consequently, to avoid storing and processing the so large-scale matrix-vector model, we use the operators $\boldsymbol{\Phi}$ and $\boldsymbol{\Phi}^\dagger$ introduced in Section II.B, in lieu of (15), to construct the model.

The MMW images of human body are relatively smooth, and the concealed weapons usually are piecewise-constant objects for which the discrete gradient turns out sparse. This property invites the applications of total variation (TV) compressive sensing techniques [38]. Since there are also very small but lethal objects, such as razor and small lighter, which can be considered as point targets, we utilize the following unconstrained optimization model to reconstruct the image.

$$\hat{\mathbf{G}} = \arg\min \left\{ \|\boldsymbol{\Phi}\{\mathbf{G}\} - \mathbf{S}\|_2^2 + \lambda_1 \|\mathbf{G}\|_1 + \lambda_2 \|\mathbf{G}\|_{\text{TV}} \right\}, \tag{16}$$

where $\lambda_1$ and $\lambda_2$ provide a tradeoff between fidelity to the measurements and noise sensitivity. In this model, the symbols $\hat{\mathbf{G}}$, $\mathbf{G}$ and $\mathbf{S}$ represent the reconstructed image, the target scene, and the scattered data, respectively, all being 3-D matrices. The symbol $\boldsymbol{\Phi}$ denotes the sensing operator, as demonstrated in (13). The $\ell_2$ and $\ell_1$ norm in (16) are given as:

$\|\mathbf{X}\|_2 = \left(\sum_i |x_i|^2\right)^{1/2}$ and $\|\mathbf{X}\|_1 = \sum_i |x_i|$, respectively, where $i = 1, 2, \cdots, N_1 \times N_2 \times N_3$ for the size of $\mathbf{X}$ being $N_1 \times N_2 \times N_3$. The TV norm is obtained by,

$$\|\mathbf{X}\|_{\text{TV}} = \\ \sum_{n_1=1}^{N_1-1}\sum_{n_2=1}^{N_2-1}\sum_{n_3=1}^{N_3-1} \left\{ \begin{array}{l} |x_{n_1,n_2,n_3} - x_{n_1+1,n_2,n_3}| + |x_{n_1,n_2,n_3} - x_{n_1,n_2+1,n_3}| \\ + |x_{n_1,n_2,n_3} - x_{n_1,n_2,n_3+1}| \end{array} \right\} \\ + \sum_{n_1=1}^{N_1-1} |x_{n_1,N_2,N_3} - x_{n_1+1,N_2,N_3}| + \sum_{n_2=1}^{N_2-1} |x_{N_1,n_2,N_3} - x_{N_1,n_2+1,N_3}| \\ + \sum_{n_3=1}^{N_3-1} |x_{N_1,N_2,n_3} - x_{N_1,N_2,n_3+1}|. \tag{17}$$

Many optimization algorithms, such as the iterative shrinkage-thresholding based algorithms [39], Least Absolute Shrinkage and Selection Operator (LASSO) [40], can be used to solve the underdetermined equation (16). Because it is not the emphasis of this paper, we just adopt the algorithm in [41] to minimize (16), which is based on the conjugate gradient method. The gradient of $f = \|\mathbf{\Phi}\{\mathbf{G}\}-\mathbf{S}\|_2^2 + \lambda_1 \|\mathbf{G}\|_1 + \lambda_2 \|\mathbf{G}\|_{TV}$ is given by,

$$\nabla f = 2\mathbf{\Phi}^\dagger\{\mathbf{\Phi}\{\mathbf{G}\}-\mathbf{S}\} + \lambda_1 \nabla\|\mathbf{G}\|_1 + \lambda_2 \nabla\|\mathbf{G}\|_{TV}. \qquad (18)$$

The non-smooth functions in (18), i.e., the $\ell_1$ norm and the TV norm, can be smoothed by using an approximate function $|x| \approx \sqrt{x^*x+\nu}$, where $\nu$ is a small positive smoothing parameter. This enables the corresponding gradient to be calculated. The reader can refer to [41].

Another key factor in CS framework is the restricted isometry property (RIP) [12], which is widely used for analyzing the performance of sparse reconstruction algorithms. However, the RIP is often difficult. An alternative to RIP is the mutual coherence of the sensing matrix, which is a more practical approach for evaluating the CS recovery properties. The mutual coherence of a sensing matrix is defined as follows [16]:

$$\mu = \mu(\mathbf{\Phi}) = \max_{i \neq j} \frac{|\langle \varphi_i, \varphi_j \rangle|}{\|\varphi_i\|_2 \|\varphi_j\|_2}, \qquad (19)$$

where $\varphi_i$ represents the $i$th column of the matrix $\mathbf{\Phi}$. If $\mu$ is small, we state that $\mathbf{\Phi}$ is incoherent. However, in this paper, we cannot directly use (19) to evaluate the mutual coherence due to the fact that we represent $\mathbf{\Phi}$ as a sensing operator as shown in (13). Instead, as has been shown in [41] and [16], the point spread function (PSF) could be used to measure the mutual coherence of a sampling scheme. It is defined as follows:

$$\text{PSF}(j,i) = \frac{\langle \mathbf{\Phi}\mathbf{e}_i, \mathbf{\Phi}\mathbf{e}_j \rangle}{\|\mathbf{\Phi}\mathbf{e}_i\|_2 \|\mathbf{\Phi}\mathbf{e}_j\|_2}, \qquad (20)$$

where $\mathbf{e}_i$ is the $i$th vector of the natural basis having 1 at the $i$th location and zeros elsewhere. The inner product of $\langle \mathbf{\Phi}\mathbf{e}_i, \mathbf{\Phi}\mathbf{e}_j \rangle = \mathbf{e}_j^H \mathbf{\Phi}^H \mathbf{\Phi} \mathbf{e}_i$ amounts to the selection of the $(i,j)$-th element of $\mathbf{\Phi}^H\mathbf{\Phi}$, which is exactly the inner product $\langle \varphi_i, \varphi_j \rangle$ in (19) if $\mathbf{\Phi}$ has a matrix form, where the superscript "H" represents the Hermitian operator. Accordingly, a simple measure to evaluate the incoherence is the maximum of the sidelobes of PSF:

$$\mu = \max_{i \neq j} \left| \frac{\text{PSF}(i,j)}{\text{PSF}(i,i)} \right|. \qquad (21)$$

Although, with the proposed approach, the mutual coherence in (19) cannot be directly computed, the PSF can be readily obtained. It is desirable to have the $\text{PSF}(i,j)|_{i \neq j}$ to be as small as possible, and have random noise-like statistics for the random undersampling schemes. From (20), we can represent $\mathbf{e}_i$ as a target scene with only one point target located at its $i$th element. Then, the imaging procedure is given by $\hat{\mathbf{e}}_i = \mathbf{\Phi}^\dagger\{\mathbf{y}\}$, where $\mathbf{y} = \mathbf{\Phi}\{\mathbf{e}_i\}$. Because $\mathbf{e}_i$ only has one nonzero element at the $i$th position, then $\hat{\mathbf{e}}_i = \mathbf{\Phi}^\dagger\{\mathbf{\Phi}\{\mathbf{e}_i\}\}$ corresponds to the $i$th column of $\mathbf{\Phi}^H\mathbf{\Phi}$ when $\mathbf{\Phi}$ is a matrix, as illustrated in Fig. 2. Note that all elements, except for those on the diagonal of $\mathbf{\Phi}^H\mathbf{\Phi}$, are sidelobes of the PSFs. The diagonal elements are the main-lobes of PSFs corresponding to the point targets located at all the possible positions in the entire target scene. In this respect, the maximum sidelobe value of $\hat{\mathbf{e}}_i$ could be used to evaluate the mutual coherence.

Based on the above discussion, we analyze the effect of different undersampling schemes on the PSF. A uniform-random undersampling scheme is applied to choose the 2-D antenna positions. This approach has been used in [42] for stepped frequency waveform design, and yield a better performance than a totally random undersampling strategy. Specifically, in uniform-random undersampling, we first divide the 2-D antenna positions into a number of non-overlapping groups. Each group consists of the same number of contiguous antenna positions out of which few are selected. In so doing, we provide more uniform illumination of the target scene. In the simulation, we perform 200 independent runs with respect to different selections of the antenna positions according to their corresponding undersampling schemes. The mean of the PSFs is computed. We project the maximum values of the mean PSFs to one dimension, such as the azimuth dimension, as shown in Fig. 3, for both random and uniform-random undersampling schemes. The level of the horizontal line measures the maximal value of the sidelobes and aliasing artifacts. It is evident from Fig. 3 that the artifacts introduced by uniform-random undersampling are lower than those by totally random undersampling. Moreover, the variation of the artifacts of the uniform-random scheme is also lower than that of the totally random scheme. In Fig. 4, we compare the PSFs for different sensing operators, such as the $\omega K$ based sensing operator and the proposed operator. It is evident that the level and variation of the artifacts and sidelobes for the proposed sensing operator are both lower than those for the $\omega K$ based operator, which indicates better incoherence. The aliasing artifacts can be removed by using a nonlinear reconstruction technique improving sparsity as introduced in [16].

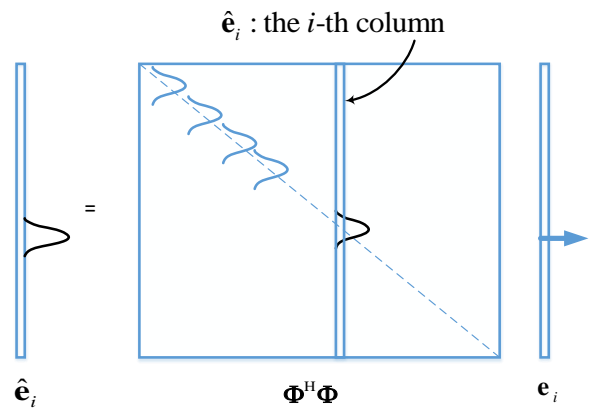

Fig. 2. Illustration of the relationship between vector $\hat{\mathbf{e}}_i$ and matrix $\mathbf{\Phi}^H\mathbf{\Phi}$.

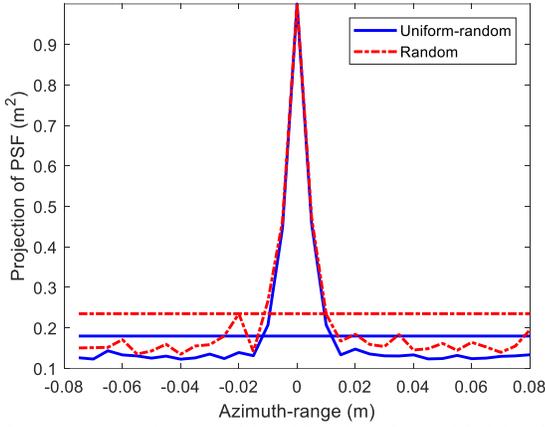

Fig. 3. Maximum projections of the PSFs using 12.5% of full data for the random undersampling scheme and the uniform-random undersampling scheme.

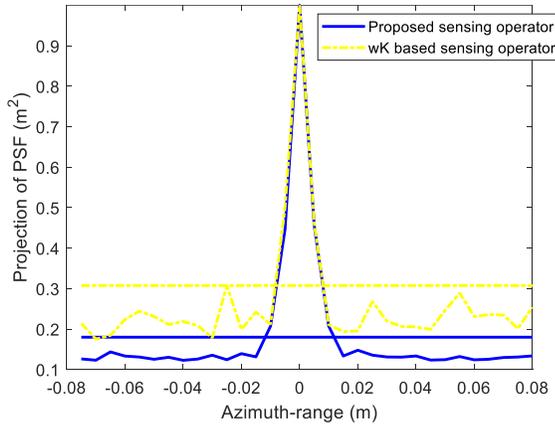

Fig. 4. Maximum projections of the PSFs using 12.5% of full data for the $\omega K$ based operator and the proposed operator, both utilizing the uniform-random undersampling scheme.

## IV. SIMULATIONS AND EXPERIMENTAL RESULTS

This section demonstrates the performance of the proposed 3-D imaging technique using simulation and real data as well as evaluates the technique computational efficiency.

### A. Simulation results

First, we compare the imaging results of the 3-D holographic algorithm with the $\omega K$ algorithm. The operating frequencies varies from 72 GHz to 76 GHz. The size of the antenna array is $64 \times 64$ elements, and the spacing of the antenna elements in both dimensions is 3 millimeter. The target model is shown in Fig. 5. Fig. 6 depicts the 3-D imaging results of both the proposed algorithm and the $\omega K$ algorithm. In order to reveal image details, we project the 3-D image onto a 2-D planes by using a maximum value projection, as shown in Figs. 7 and 8, which, respectively, correspond to the azimuth-range vs. elevation-range and the azimuth-range vs. down-range projections. Both Figures are plotted with a dynamic range of 30 dB. It is noted that there are no visible differences between the imaging results of the two algorithms, and both obtain a high resolution image of the target.

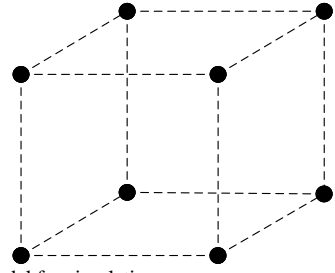

Fig. 5. Target model for simulation.

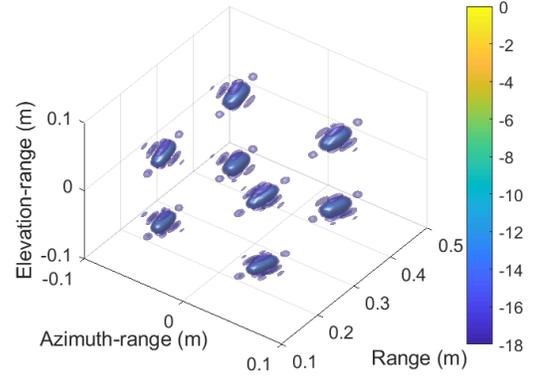

(a)

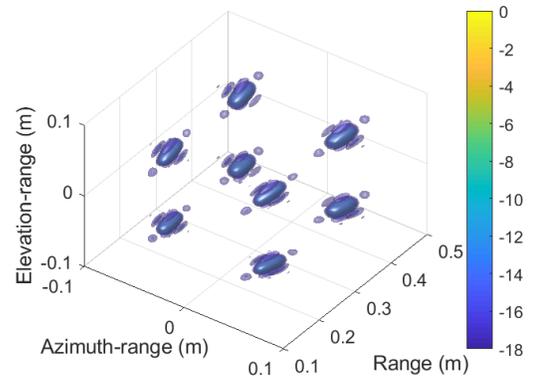

(b)

Fig. 6. 3-D plot of the imaging results. (a) the 3-D holographic algorithm; (b) the $\omega K$ algorithm.

Next, we construct the sensing operators for CS optimization model for both algorithms, as described in Sec. III. We present the computational time for the Holography-CS and the $\omega K$-CS in Fig. 9. The size of the reconstructed image varies from $16 \times 16 \times 16$, $16 \times 32 \times 32$ to $16 \times 64 \times 64$. A Dell OPTIPLEX 7010 desktop computer with four Intel Core i7 processors @3.4GHz and 16-GB memory is used for the simulations. Clearly, the Holography-CS method is much faster than the $\omega K$-CS method. The computational time of the $\omega K$-CS could be ten times of that of Holography-CS. The difference becomes even more pronounced if we utilize the fact that the imaging region along the range direction can be selected for the Holography-CS, and we can just reconstruct the image over only the region of interest. On the other hand, the $\omega K$-CS method needs to reconstruct the image over the entire unambiguous range which is determined by the sampling interval of the operating frequencies.

Fig. 10 illustrates the normalized root mean square errors (RMSEs) of the above two operator-based CS methods as well as those of the traditional Fourier based method (specifically, the $\omega K$ algorithm), for different signal-to-noise ratios (SNRs) and data ratios. The RMSE can be calculated by $\text{RMSE} = \sqrt{(1/N_R N_A N_E) \sum_{n=1}^{N_R} \sum_{p=1}^{N_A} \sum_{q=1}^{N_E} \left[ \mathbf{G}(n,p,q) - \hat{\mathbf{G}}(n,p,q) \right]^2}$, where $\mathbf{G}$ and $\hat{\mathbf{G}}$ represent the target model and the corresponding reconstructed image, respectively, both with the size of $N_R \times N_A \times N_E$. We perform 50 independent runs for each SNR and data ratio to obtain the error means and standard deviations. It is noted from Fig. 10 that the means and standard deviations of RMSEs of the operator-based CS methods are both much lower than those of the Fourier based method (except for the data ratio 0.2, the deviation of the Fourier method is smaller than that of the CS methods). The Holography-CS method has smaller errors than the $\omega K$-CS. On the other hand, the errors of these two methods exhibit a small increase when the data ratio exceeds some extent. It could be caused by the combined effects of mutual coherence and data ratio.

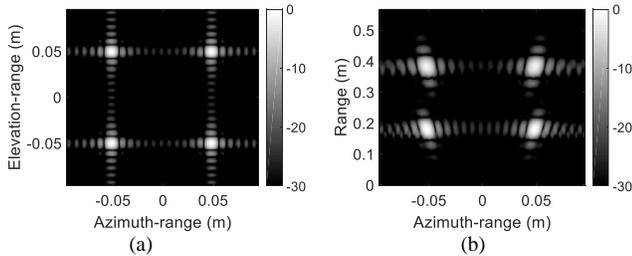

Fig. 7. Maximum projected imaging results of the 3-D holographic algorithm. (a) azimuth-elevation dimensions; (b) azimuth-range dimensions.

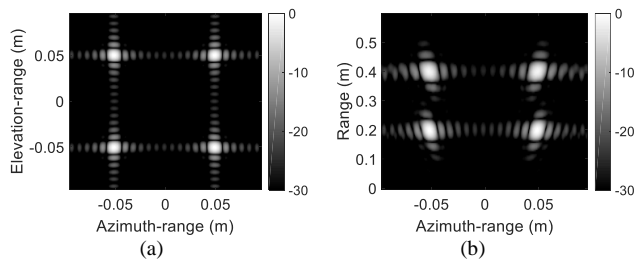

Fig. 8. Maximum projected imaging results of the $\omega K$ algorithm. (a) azimuth-elevation dimensions; (b) azimuth-range dimensions.

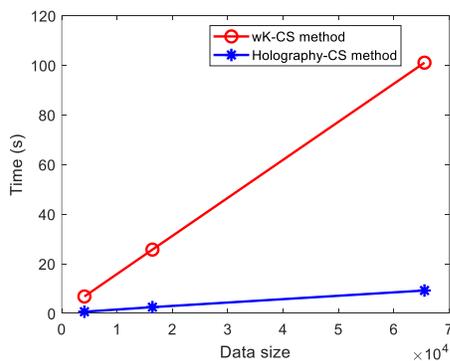

Fig. 9. Comparison of the computational time with different image size.

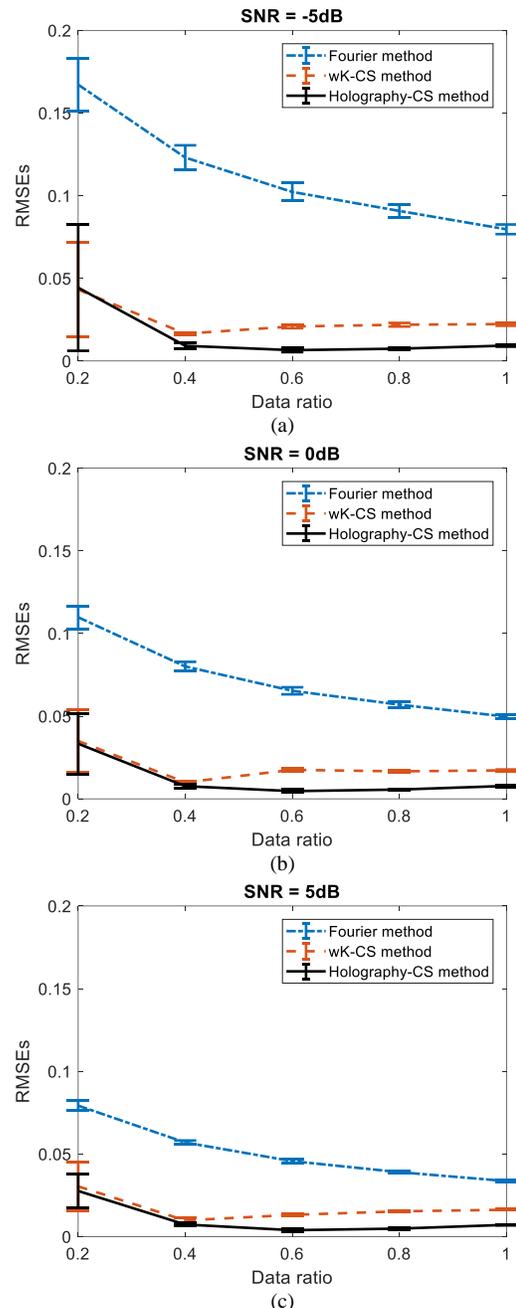

Fig. 10. Comparison of the RMSEs with respect to different SNRs.

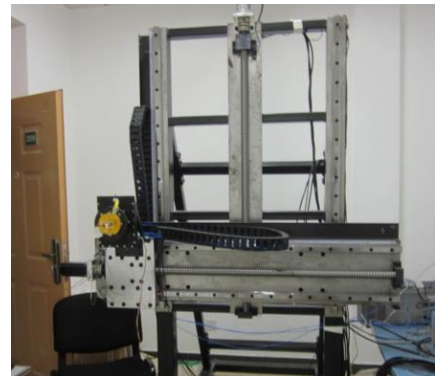

Fig. 11. W-band imaging system.

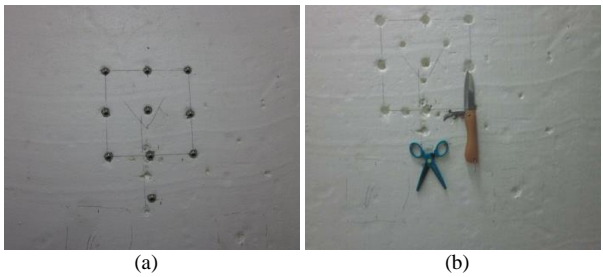

Fig. 12. Imaging targets; (a) ten metal balls, (b) knife and scissor.

### B. Experimental results

For the real measured data, we provide a comparison between the Holography-CS and Fourier-based imaging method. The results of $\omega K$-CS method are not given because its computational time is extremely long. We construct a W-band (92GHz to 94 GHz) imaging system, as shown in Fig. 11. The antenna can be scanned at a 2-D planar aperture with $200 \times 200$ points.

The imaging targets include ten small metal balls and a combination of knife and scissor, representing a point-like target and volume target, respectively, as illustrated in Fig. 12. The imaging results of Fourier-based method (specifically, the $\omega K$ algorithm) are shown in Figs. 13 and 14, by using the full data set, 50%, and 30% of the data, respectively. Clearly, the random undersampling results in white noise like artifacts in the image domain. These artifacts cannot be removed by the Fourier-based methods. However, with CS, it is possible to remove the aliasing noisy artifacts without degrading the image quality. Figs. 15 and 16 illustrate the imaging results of the Holography-CS method. It is evident that the CS-based method can obtain better imaging results than the traditional Fourier-based method, even when using much reduced data.

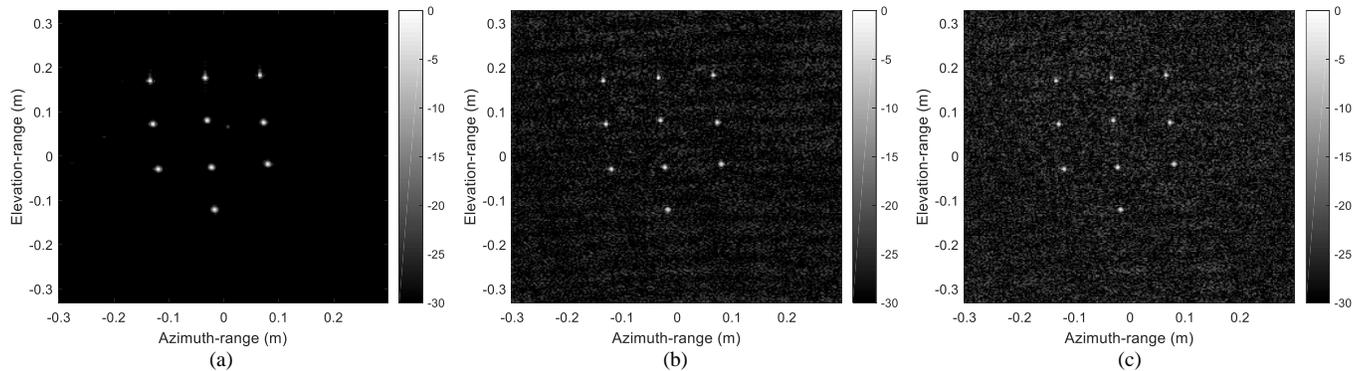

Fig. 13. Fourier-based imaging results with respect to different data ratios; (a) full data, (b) 50% of full data, (c) 30% of full data.

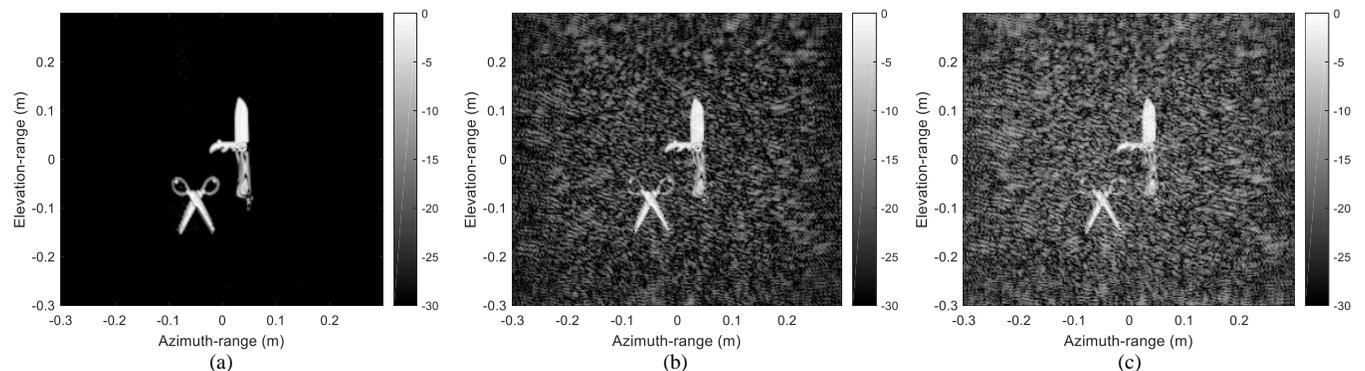

Fig. 14. Fourier-based imaging results with respect to different data ratios; (a) full data, (b) 50% of full data, (c) 30% of full data.

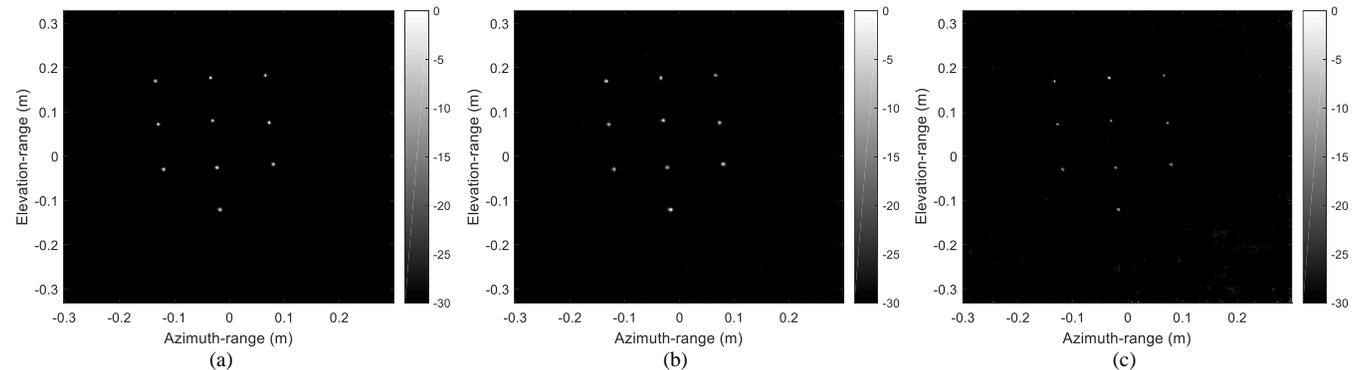

Fig. 15. Holography-CS imaging results with respect to different data ratios; (a) full data, (b) 50% of full data, (c) 30% of full data.

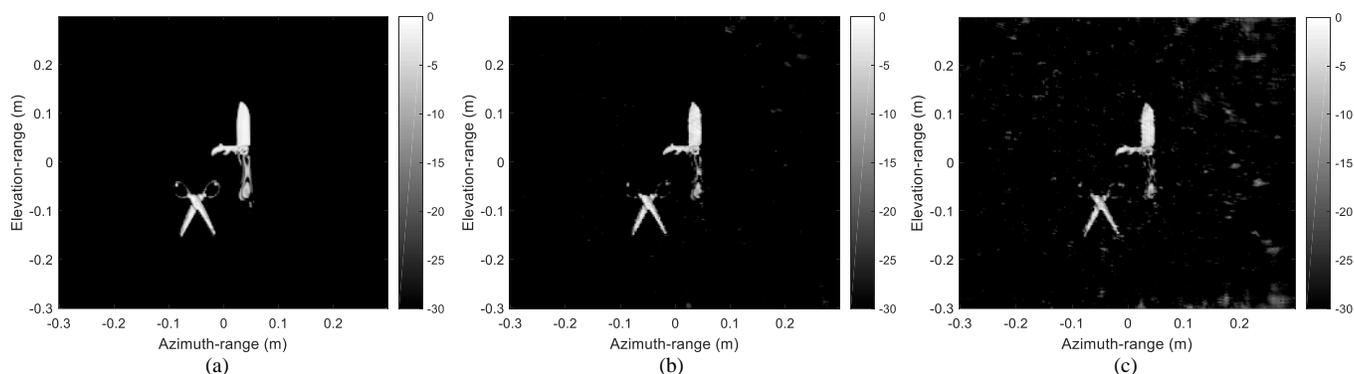

Fig. 16. Holography-CS imaging results with respect to different data ratios; (a) full data, (b) 50% of full data, (c) 30% of full data.

## V. CONCLUSIONS

This paper proposed a 3-D CS method to the near-field MMW imaging. Based on a 3-D holographic imaging algorithm, we constructed a sensing operator to avoid storing and processing of the large-scale sensing matrix. Most importantly, there are no interpolations for both the forward and backward operators in performing the optimization procedure iterations. We discussed the mutual coherence and PSF when dealing with sensing operator, and suggested a semi-random way of compressing the antenna elements. The proposed 3-D imaging technique has less computations, better performance and improved imaging quality compared to the $\omega K$ based CS imaging algorithm. Simulations and experimental results also demonstrated that the proposed technique, not only improves over the Fourier-based imaging, but also outperforms the $\omega K$ based CS method with lower RMSEs.